\documentclass[twocolumn,showpacs,preprintnumbers,amsmath]{revtex4}
\usepackage{graphicx}
\usepackage{dcolumn}
\usepackage{bm}
\begin{document}

\title{Terahertz wave emission from mesoscopic crystals of BSCCO}
\author{Wei Zhou$^{1,2}$, Chen Wang$^1$, and Qing-Hu
Chen$^{2,1,*}$ }
 \affiliation{ $^{1}$ Department of Physics, Zhejiang
University,
Hangzhou 310027, P. R. China  \\
$^{2}$ Center for Statistical and Theoretical Condensed Matter
Physics, Zhejiang Normal University, Jinhua 321004, P. R. China }
\date{\today}
\begin{abstract}
 We study the Terahertz (THz) emission mechanism in BSCCO with an external
magnetic field theoretically. The total power is proportional to the
square of the number of layers and the frequency. Under the small
size limitation, the angular distribution is similar to that in the
dipole emission. The present theoretical results for THz power
output are in the same order of magnitude as the experimental ones,
in contrast with previous simulations. A series of size-dependent
intrinsic frequencies are also observed. When the external frequency
induced by the external current resonates with intrinsic ones, the
distinct THz emission is observed.

\end{abstract}


\maketitle

Highly anisotropic high-Tc superconductor
Bi$_2$Sr$_2$CaCu$_2$O$_8$(BSCCO) is considered as a stack of
intrinsic Josephson junctions(IJJs) on atomic scale
\cite{Kleiner,Wang}. This materials may provide a new way to
generate terahertz (THz) electromagnetic(EM) wave and therefore
possibly fills the "THz gap"\cite{Gap,OZYUZER,T134515,
Bae027002,LinB,Savelev,Koyama2009}, one area of the EM spectrum
between microwave and infrared frequencies, which has not yet been
exploited. The THz EM-wave emission based on AC Josephson effect of
IJJs has been detected by Ozyuzer {\itshape et al}. \cite{OZYUZER},
and  also studied extensively in the theoretical side. Some
mechanisms of the THz emission have been proposed\cite{T134515,LinB}
recently. However, these mechanisms can not bridge  the big gap for
the power in simulations\cite{LinB} and
experiments\cite{OZYUZER,Bae027002}.

In this paper, we  successfully develop a new boundary conditions,
which explains the dipole angular distribution and predicts  that
the power is proportional to the square of layer number and
frequency. By solving the sine-Gordon equations numerically with a
proper boundary conditions, we observe  a stable power output in
some parameter regime. The present power density is in the same
order of magnitude as the experimental results. Finally, the detail
mechanism is  suggested.

The plasma frequency of single Josephson junction is in subterahertz
wave band. There is a collective cavity resonance mode in the stack
of IJJs, which is crucial to enhance the power and to increase the
frequency from sub-THz to THz. IJJ is formed naturally in BSCCO
where Bi-Sr-O layers between the superconducting CuO$_2$ layers acts
as a non-conducting barrier. It can be described by the well known
coupled sine-Gordon equation\cite{Kleiner}
\begin{eqnarray}
\frac{{\partial ^2 \phi_{l + 1,l} }}{{\partial x^{'2} }}&=&(1 - \zeta
\Delta^{(2)} )\{ \partial _{t' }^2 \phi_{l + 1,l}  + \beta
\partial _{t' } \phi_{l + 1,l}
\nonumber\\&&
+ \sin \left( {\phi_{l + 1,l} } \right) \}.
\label{eq:sine}
\end{eqnarray}
%\begin{eqnarray}
%(1-\alpha\Delta^{(2)})
%\left( {\frac{{\partial ^2 \phi_{l + 1,l} }}{{\partial x^{'2} }} +
%\frac{{\partial ^2 \phi_{l + 1,l} }}{{\partial y^{'2}}}}\right)=
%(1-\zeta\Delta ^{(2)})
%\nonumber\\
%\times\left\{ { \partial _{t'}^2 \phi_{l + 1,l} +
%\beta \partial _{t' } \phi_{l + 1,l} + (1 - \alpha \Delta ^{(2)})
%\sin \phi_{l + 1,l} } \right\},
%\label{eq:tachiki1}
%\end{eqnarray}
where $x'= x/\lambda _c$ and $y'=x/\lambda _c$ are the dimensionless
positions in the a-b plane,  $t'=\omega _p t$ is the dimensionless
time  with $\omega _p=c/\lambda _c\sqrt{\varepsilon _c }$   the
Josephson plasma frequency. The current interlayer coupling
parameter $\zeta$ is described as $\lambda _{ab}^2/sD$, where $s$
and $D$ are the superconducting and insulating layer thickness.
$\beta  =4\pi \sigma _c \lambda _c/c\sqrt {\varepsilon _c }$ is the
resistive dissipation factor, where ${\varepsilon _c }$ and ${\sigma
_c}$ are the dielectric constant and the conductivity of the
insulating layers, respectively. The abbreviation operator of
$\Delta ^{(2)}$ is defined as $\Delta ^{(2)} f(l) = f(l + 1) - 2f(l)
+ f(l - 1)$.

%%%%%%%%%%%%%%%     BOUNDARY CONDITION     %%%%%%%%%%%%%%%%%%%%%%%%%%%%
\begin{figure}
\includegraphics[scale=0.5]{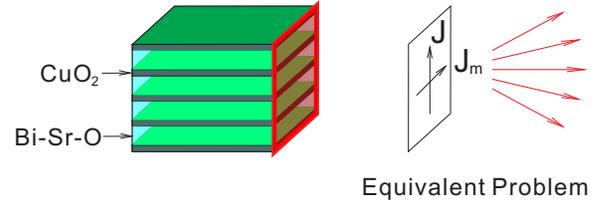}
\caption{\label{fig:fig1}(Color online).
Schematic view of BSCCO source (left) and the
equivalent radiation sources (right).
}
\end{figure}
It is known that the boundary conditions are very crucial for the
wave emission. The BSCCO source and outside media region are
separated by a boundary surface. On the interface, the surface
integrals of $E$ and $H$ are constant, except in the red region
shown in the Fig. \ref{fig:fig1}.  It is an effective radiation
section where $H$ parallel to y-axis and $E$ to  x-axis are
determined through $\phi_t=2eDE_z/\hbar$ and $\phi_x=2\pi \mu _0
DH_y/\Phi _0 $  respectively. So the EM-wave is described by Maxwell
equations outside the BSCCO region and sine-Gordon equations inside
the BSCCO region. To evaluate the far field, the original problem
could be transformed to an equivalent one described  by Love's
Formulation\cite{LOVE}.  In this way,  the original IJJ region is
replaced by a free space, the BSCCO source is removed, and
equivalent electric and magnetic currents given below are suspended
on the boundary surface,
\begin{equation}
J = n \times H,
\nonumber
\end{equation}
\begin{equation}
J_m  =  - n \times E,
\nonumber
\end{equation}
where $n$ is an  unit vector normal to the effective radiation
section,   $J$ and $J_m$ are the equivalent sources producing the
same $E$ and $H$ in the region outside. According to the
superposition principle of the electromagnetic waves, the $E$ and
$H$ in far field are the sum of contributions from electric sources
and magnetic sources respectively. The angular distribution of total
power is determined by the radiation from both  $J$  and $J_m$. For
the electric sources, the vector potential in outside region is
generated by $J$,
\begin{equation}
A(r)=\frac{\mu_0}{4\pi}
\int {dS\int{d\omega}}
\frac{J(r',\omega )e^{ik|r - r'| - \omega t}}{|r - r'|}.
\label{eq:A0}
\end{equation}
The average energy flow density can be evaluated by  the time
averaged Poynting vector
\begin{equation}
\bar S(r)=
\frac{L_y^2N_J^2D^2 \mu _0}{32\pi ^2 c}
\int {d\omega }
\frac{\omega ^2 J^2 (\omega )}{r^2}\sin ^2 \theta.
\label{eq:S}
\end{equation}
It is Eq.(\ref{eq:S}) that can account for a lot of  experimental
and numerical observations, such as the angular distribution of the
emission,  total power which is proportional to $N_J^2$, and
considerable power discrepancies  between  simulations and
experiments.

%%%%%%%%%%%%%%%%%%%%    ANGULAR DISTRIBUTION     %%%%%%%%%%%%%%%%%%%%%%
\begin{figure}
\includegraphics[scale=0.5]{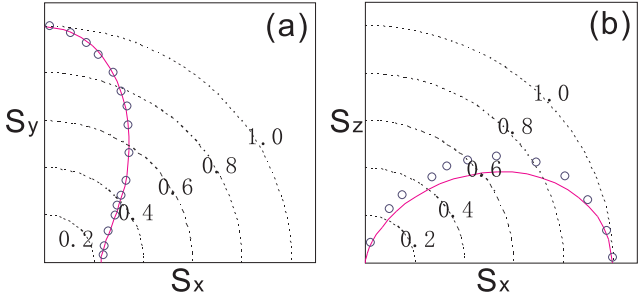}
\caption{\label{fig:fig2}(Color online). (a) Angular distribution of
the radiation energy in ab-plane. The present analytical results are
denoted by the solid curve and  the previous simulation ones by the
open circles. (b) Angular distribution of the radiation energy in
ac-plane.}
\end{figure}

The analytical results for the angular distribution of the radiation
energy by Eq. (3) is plotted in Fig. \ref{fig:fig2} with the solid
curves. Fig. \ref{fig:fig2}(a) shows the results of the xy model
with $L_x=300\mu m, L_y=100\mu m$. This model describes the in-phase
EM emission from IJJs with infinite junction number. The angular
distribution of the $J_m$ has the  cylindrical symmetry, and The
angular dependence  of $J$ can be derived from Eq. (\ref{eq:A0})
\begin{equation}
S_J \propto \left[ {\frac{{\sin \theta \cos \left( {\frac{\pi }{2}\gamma \cos
\theta }\right)}}{{\left( {\gamma \cos \theta } \right)^2  - 1}}} \right]^2 ,
\end{equation}
where the $\gamma  = \lambda _h /\lambda =1.36$. $\lambda _h$ is the
standing wave length of $E$ at the boundary, and $\lambda$ is the
wave length of the far field EM wave. Fig. \ref{fig:fig2}(b) shows
the radiation in xz-plane of the  sample with  $L_x=100\mu m$ and
$L_z=0.2\lambda_c$. The thickness of the source is much smaller than
the wave length. Neglecting the $r'$ in Eq. (\ref{eq:A0}), the
Poynting vector is proportional to $\sin^2\theta$, the dipole
angular distribution. It is the small size of effective radiation
section that generates the dipole angular distribution.

By using a 2D model, Koyama {\itshape et al} \cite{Koyama2009}.
simulated the angular distribution of the radiation energy. Their
numerical results are also list in Figs. \ref{fig:fig2}(a) and (b)
by  open circles. It is very interesting that our analytical results
agrees quite well with the simulation ones.

%%%%%%%%%%%%%%%%%%%%%%%%         POWER           %%%%%%%%%%%%%%%%%%%%%%%%
It was recently observed that small samples can be used  to reduce
the heating effect, therefore,  the small size approach has been
generally adopted. When the EM radiation from the IJJs is
approximately monochromatic in THz band, the total power is
\begin{equation}
P=
\frac{L_y^2N_J^2\omega ^2\Phi _0^2}{96\pi^3 c}
\left[
\frac{\phi_x^2}{\mu _0}
+\varepsilon_0 \phi _t^2
\right].
\label{eq:power}
\end{equation}
When the change of the cavity resonance is not obvious,
$\phi_x(\omega)$ and $\phi _t(\omega)$ is constant. Thus $S_{eff}^2
\omega^2/P$ almost  remains unchanged, where $S_{eff}$ is the area
of the effective section. The total power is proportional
to the square of the number of layers, consistent with the
experimental observations\cite{OZYUZER}.

\begin{figure}
\includegraphics[scale=0.5]{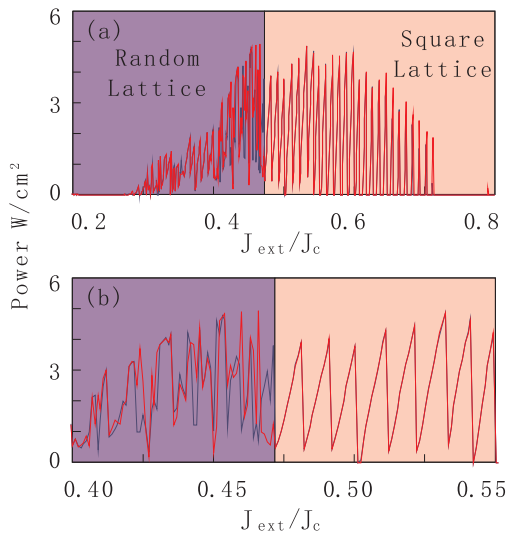}
\caption{\label{fig:power} (Color online). Radiation power as a
function of the external current. The blue curve represents the
radiation power averaged from 13.67 ns to 27.34 ns, the red curve
the radiation power averaged from 27.34 ns to 41.01 ns. (a) In the
pink regime, the vortices are distributed almost randomly. In the
purple regime, the moving Josephson vortices form the in-phase
rectangular lattice. (b) is an enlarged view of (a).}
\end{figure}

The power of radiation is also simulated with the uses of the
sine-Gordon equations and Eq. (\ref{eq:power}). Since the magnetic
field of the THz EM wave is much smaller than the external one, the
boundary condition can be set as  $\phi_{x'}=H',
H'=2\lambda_{c}DH$\cite{Cirillo}. For such a boundary condition, the
Poynting vector can not be calculated directly\cite{Bulaevskii}, but
Eq. (\ref{eq:power}) provides another way to the evaluation of  the
power. The parameters used here are $\lambda _{ab}  = 0.4\mu m$,
$\lambda_{c}=200\mu m$, $s=0.3nm$, $D=1.2nm$, $\beta=0.02$,
$L_x=20\mu m$ and $L_y=20\mu m$. The number of junctions is
$N_J=30$. The power densities obtained in the  previous
simulations\cite{LinB,T134515} are independent of the effective
section area. Lin {\itshape et al}.\cite{LinB} estimated the power
density to be around $400 W/cm^2$, and Tachiki {\itshape et
al}.\cite{T134515} about 3000 W/cm$^2$. In sharp contrast with the
previous reports, Eq. (\ref{eq:power}) shows that the power density
is proportional to $S_{eff}$. Our effective section area is set to
$0.9 \mu m^2$, close to the experimental value\cite{Bae027002}. Fig.
\ref{fig:power} shows the external current dependence of the power
intensity for the external magnetic field  $H_{ext}=2T$ along the y
axis. The blue curve denotes the  power averaged from 13.67 ns to
27.34 ns, and the red one  from 27.34 ns to 41.01 ns. Because of the
appropriate boundary conditions we employed, our result for the
power density is much smaller than previous numerical ones reported
in the literature\cite{LinB,T134515}. The maximum output for the
power density is about 6 W/cm$^2$, which is just in the same order
of magnitude as the experimental result observed  by Bae {\itshape
et al}.\cite{Bae027002}.

Both Figs. \ref{fig:power}(a) and (b) have been divided into two
different regimes, marked with different background colors. When
$J_{ext}<0.47J_c$ (in the purple regime), the vortices are randomly
distributed. In general, the power roughly increases  with
$J_{ext}$. And if we look into details, it could be found that the
blue line and red line do not coincide, due to an unstable power
caused by random flux flow. When $J_{ext}>0.47J_c$, the vortex
structure is transformed into the in-phase rectangular lattice, a
stable configuration. These configurations are discussed
analytically by Koshelev\cite{Koshelev}. In this situation, the
distinct power radiation appears. In addition, the output power
fluctuates within the pink regime. This phenomenon is closely
related to the radiation mechanism, and will be discussed below.

It is interesting to note that only a small part  of the input
power, not more than 5\%, is able to emit out as EM-wave in all
previous experiments \cite{OZYUZER,Bae027002}. Eq. (\ref{eq:power})
can give a reasonable explanation to theses experimental findings.
When $L_yN_J$ is small, the interior surfaces reflect a wave of low
frequencies. The inner EM-wave, generated by AC Josephson effect,
bounces back and forth within BSCCO. Like a cavity resonator, where
the EM field forms standing waves, the EM waves with resonance
frequencies can also form in BSCCO. Every discrete peak in Fig.
\ref{fig:power} provides an evidence of the standing waves.  Fig.
\ref{fig:power}(b) is an enlarged view of Fig. \ref{fig:power}(a).
It is clear that the distances between neighboring peaks are almost
the same. These peaks just correspond to the characteristic
frequencies with different resonance modes.

%%%%%%%%%%%%%%%%%%%%%%%%%%%%%%%%%%%%%%%%%%%%%%%%%%%%%%%%%%%%%%%
\begin{figure}
\includegraphics[scale=0.5]{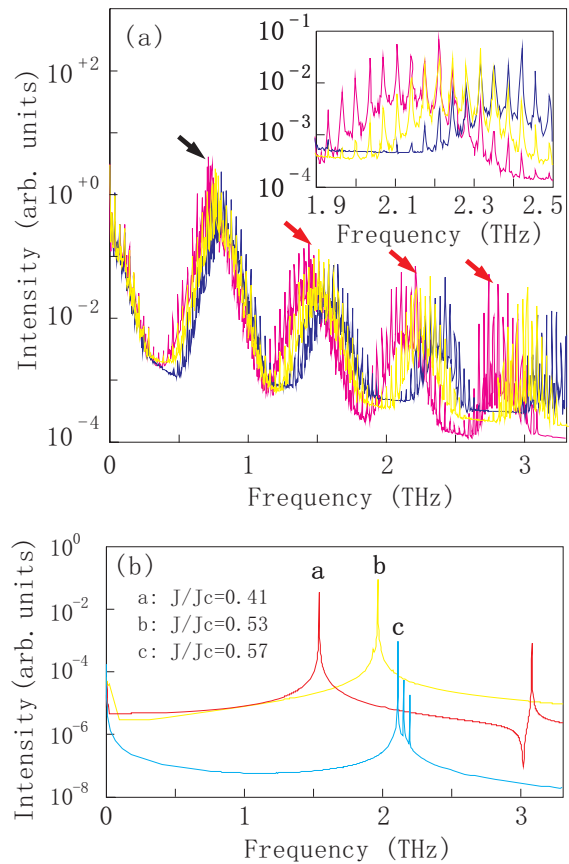}
\caption{\label{fig:freq}(Color online). (a) The frequency spectra
of the power. The external current are $0.21J_c$(pink),
$0.22Jc$(yellow) and $0.23J_c$(blue). Arrows denote the major
frequencies. The inset shows the enlarged view. (b) Frequency
spectra of the power with high $J_{ext}$. }
\end{figure}

The different flux-line lattice structures results in  the different
V-I curves and frequency characteristics \cite{Machida}. To
understand the resonance mechanism, we analyze the frequency spectra
with different external currents. As plotted in the Fig.
\ref{fig:freq}(a) and (b), the frequency spectra of random lattice
is much different from that of the rectangular lattice.

Fig. \ref{fig:freq}(a) shows the frequency spectra with external
currents $J_{ext}=0.21J_c$, $0.22J_c$ and $0.23J_c$. Under these
conditions, the output power is less than 1\% of the maximum. With
low $J_{ext}$, the spectra are not monochromatic and lots of peaks
appear. The envelope lines of these peaks have characteristic
frequencies (black arrow) and higher harmonics (red arrows). These
frequencies are generated from the $\sin (\phi )$. Integration of
sine-Gordon equations gives $\langle{\beta\phi
_t}\rangle+\langle{\sin(\phi)}\rangle=J_{ext}$, where
$\langle{\beta\phi _t}\rangle$ is the normal current and
$\langle{\sin(\phi)}\rangle$ is the supercurrent. As
$\langle{\beta\phi _t}\rangle$ is not zero, the base frequency of
$\sin (\phi )$ is derived
\begin{equation}
\frac{\langle{\phi_t}\rangle}{2\pi}=\frac{J_{ext}-\langle{J_s}\rangle}{2\pi\beta}.
\label{eq:jf}
\end{equation}
The frequency, based on the external current,
is of great importance to tune the supercurrent frequency.
Furthermore, it can be used to control the radiation frequency and power.

Fig. \ref{fig:freq}(a) shows a discrete cavity resonance and an
equidistant frequency difference. The phenomenon can be clearly seen
in the inset, which is an enlarged view. It is also interesting to
note that in the three lines for different external currents
$J_{ext}=$ $0.21J_c$(pink), $0.22Jc$(yellow) and $0.23J_c$(blue),
the peaks are in the same frequencies. It follows  that these
frequencies are intrinsic and independent of the external
parameters, such as the external current, the magnetic field, and so
on.  The differences ($\Delta\omega$) between two neighboring
frequencies could be regarded as the characteristic frequencies.

\begin{figure}
\includegraphics[scale=0.5]{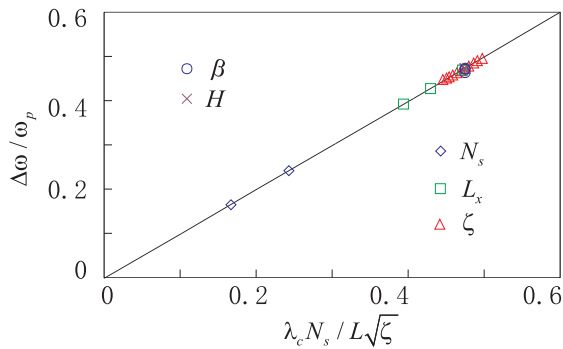}
\caption{\label{fig:fig5}(Color online). The intervals of the
intrinsic frequencies. Only one of the parameters($N_s$, $H$, $L_x$,
$\zeta$ and $\beta$) is changed in same group of data points.}
\end{figure}

In order to clarify these intrinsic frequencies, $\Delta\omega$ is
calculated under different $J_{ext}$, $H_{ext}$, length, layer
number, and coupling $\zeta$. The results  are divided into 5 groups
and collected in Fig. \ref{fig:fig5}. The parameters for the same
group of data are fixed, except that one parameter denoted in the
legend of Fig. \ref{fig:fig5} can be tuned.   Because the
frequencies are independent of $\beta$ and $H$, the open circles and
crosses with different $\beta$ and $H$ coincide. From this figure,
we can observe that all data point can collapse onto a straight line
described by  $\Delta\omega= \frac{N_s c}{L_x\sqrt{\zeta\varepsilon
_r}}$. It follows that the intrinsic frequencies is $\frac{n_x N_s c
}{L_x \sqrt{\zeta\varepsilon _r}}$, where $N_s=N_J+1$ is the number
of CuO layers  and $n_x$ is the natural number. Through observing
the simulation data of the inner EM field, it could be found that
$n_x$ is the number of the standing wave nodes. On possible
explanation for these phenomena is the EM resonance between z and x
directions. The operator $(1-\zeta\Delta^{(2)})$ in Eq.
(\ref{eq:sine}) could be regarded as $(1-\partial _{z'}^2)$ with the
dimensionless quantity $z' = z_{eff} /\lambda _c$, and the effective
size $z_{eff}=z\lambda _c /\left( {s+D}\right) \sqrt \zeta$. The
resonance occurs when $\omega /k_x=\omega_z/k_{eff}$.

For high external current, although the higher harmonics sill exist,
the frequency spectra of rectangular lattice is almost
monochromatic, as shown in Fig. \ref{fig:freq}(b), close to the
experiment\cite{Kadowaki2008} . When the external frequency
estimated by Eq. (\ref{eq:jf})  equals to one of the intrinsic
frequencies, the external current would excite a monochromatic
resonance mode and the radiation would generate the maximum power,
as denoted by the curves $a$ and $b$. On the other hand, the
external current would excite several resonance modes, when it is
not equal to any intrinsic frequency, as shown by the curve $c$. The
pink regime of Fig. (\ref{fig:power}) shows these resonance peaks
with different $J_{ext}$. The distance ($\Delta J_{ext}$) between
two neighbor peaks is $0.0098J_c$, consistent with the numerical
result $0.0093J_c$ based on intrinsic frequencies.

For   fields parallel to the $ab-$plane in BSCCO, vortices  are
periodically distributed. The distributions of supercurrent and the
phase of the order parameter are also periodic. The normal current
is no longer uniform, which  results in periodical perturbation in
the inner electric field. Therefore the vortices excite the
alternative EM field. When the vortices are in motion,  the
alternative EM field is also propagating and decaying in BSCCO. In
order to form a stable alternative EM field, the vortices and the EM
field must move synchronously.

By increasing or decreasing the external current, the velocity of
the flux flow is also changed. When $J_{ext} < 0.47J_c$, the flux
flow would move more slowly than the EM field.  To achieve
synchronies along the x-axis, the normal direction of the EM
wavefront is not parallel to the flow direction. Because the
boundary surface and the wavefront is not tangential, which weakens
the power emission. On the other hand, when $J_{ext} > 0.47J_c$, the
external frequency would possibly resonate with the intrinsic
frequencies, which pumps considerable energy into the inner
alternative EM field. Since the boundary surface is tangential to
the wavefront, the power emits much more efficiently.

%%%%%%%%%%%%%%%%%%%%      ACKNOWLEDGE       %%%%%%%%%%%%%%%%%%%%%%%%%%%%%%%%%%%%%%%%%

The authors acknowledge useful discussions with X. Hu and S. Lin.
This work was supported by National Natural Science Foundation of
China under Grant No. 10774128,  PCSIRT (Grant No. IRT0754) in
University in China,  National Basic Research Program of China
(Grant Nos. 2006CB601003 and 2009CB929104).

$*$ Corresponding author. Email:qhchen@zju.edu.cn

\end{document}